\documentstyle[12pt]{article}
\def\Eq{\begin{equation}}	\def\End{\end{equation}}
\def\Eqa{\begin{eqnarray}}	\def\Enda{\end{eqnarray}}
\def\Endl#1{\label{#1} \End}	
\def\puteq#1{eq.~(\ref{#1})}	\def\ie{{\it i.e.}}  \def\eg{{\it e.g.}}
\def\ord#1{{\cal O}(#1)}	\def\d{{\rm d}}
\def\dslash{\partial\!\!\!/}	
\def\Veff{V_{\rm eff}}		\def\Fpot{F^{\rm pot}}

\input epsf
\def\boxit#1{\vbox{\hrule\hbox{\vrule{#1}\vrule}\hrule}}
\def\INSERTFIG#1#2#3{\epsfxsize=#1in \begin{center}
  \hbox to\hsize{\hfil\boxit{\epsffile{#2}}\hfil} {\sl #3} \end{center}}
\def\figskip{\newpage}

\begin{document}
\begin{titlepage}
\begin{center}
June, 1994	\hfill CMU-HEP94-20\\
{}~		\hfill DOE/ER/40682-74\\
{}~		\hfill PITT-94-06\\
{}~		\hfill hep-ph/9406422\\
\vskip1.0in {\Large\bf
Thermal Activation Rates in the Chirally Asymmetric Gross-Neveu Model}
\vskip.3in
Daniel Boyanovsky\footnote{Email: \tt boyan@vms.cis.pitt.edu}$^{(a)}$,
David E. Brahm\footnote{Email: \tt brahm@fermi.phys.cmu.edu}$^{(b)}$,\\
Richard Holman\footnote{Email: \tt holman@cmphys.phys.cmu.edu}$^{(b)}$, and
Da-Shin Lee\footnote{Email: \tt dashin@phyast.pitt.edu}$^{(a)}$\\~\\
{\it $^{(a)}$Dept.\ of Physics and Astronomy, University of Pittsburgh,
  Pittsburgh PA 15260}\\
{\it $^{(b)}$Carnegie Mellon Physics Dept., Pittsburgh PA 15213}
\end{center}

\vskip.5in
\begin{abstract}
We address the problem of how to incorporate quantum effects into the
calculation of finite-temperature decay rates for a metastable state of a
quantum field theory. To do this, we consider the Gross-Neveu model with an
explicit chiral symmetry breaking term, which allows for a metastable state.
This theory can be shown to have a ``critical bubble'' which is a solution to
the {\em exact} equations of motions (\ie\ to all orders in perturbation
theory, including all higher derivative, quantum and thermal corrections).
This configuration mediates the thermal activation of the metastable vacuum to
the true ground state, with a decay rate $\Gamma \propto \exp(-F_c/T)$, where
$F_c$ is the free energy of the critical bubble.  We then compare this exact
calculation to various approximations that have been used in previous work.
We find that these approximations all {\em overestimate} the activation rate.
Furthermore, we study the effect of finite baryon number upon the bubble
profile and the activation barriers. We find that beyond a critical baryon
number the activation barriers disappear altogether.
\end{abstract}
\end{titlepage}
\setcounter{footnote}{0}


\section{Introduction and Motivation}

Thermally activated reactions appear in many contexts, including chemical
reactions\cite{chem}, cosmological phase transitions\cite{cpt}, and
sphaleron-mediated baryon number violating transitions in the early
universe\cite{sphal}. The physically important quantity in these transitions is
the activation rate; this sets the time scale for the process, and determines
whether these reactions can be in local thermal equilibrium.

There is a standard way of computing the activation rate in a field theory at
finite temperature. Consider a field theory with a metastable false vacuum
state as well as a stable ground state. The first step in the calculation is to
construct a static, extremal configuration, corresponding to a critical bubble,
in which the field is near the true vacuum on the inside, and in the false
vacuum outside the bubble. Once this configuration is obtained, the activation
rate $\Gamma$ is proportional to $\exp(-F_c/T)$, where $T$ is the
temperature, and $F_c$ is the free energy of the critical bubble.

In most applications, the bubble configuration used in practice is a static
extremum of the {\em classical} action\cite{cc,fTref}.
 However, this leaves open the question of how to
incorporate quantum effects into the rate calculation\cite{gleis}. In
particular, it is not clear whether the {\em full} theory has a critical
bubble. Even if it does, the relation between the rate obtained by the
using the classical bubble and the rate we would calculate using the exact
bubble is not altogether clear, to say the least.

In a general field theory, it is usually impossible to construct the exact
critical bubble of the theory, so that the questions asked above are moot.
Nonetheless, it would be of great interest to have some idea of how the exact
and the approximate rates compare. It is for this reason that we consider the
Gross-Neveu\cite{gn} (GN) model in $1+1$ dimensions. In the presence of a
chiral symmetry breaking term, the theory has a metastable ground state. Of
greater use to us, however, is the fact that we can obtain the {\em exact}
critical bubble of the full quantum theory! Being able to do this allows us to
proceed with the program described above. In particular, there are {\em no}
``missing loops'', or higher derivative corrections. While this is a theory in
$1+1$ dimensions, the fact that it has much of the same richness as $3+1$
theories (\eg\ asymptotic freedom, dimensional transmutation, etc.) leads us
to think that some of the results of this paper could be applicable to
realistic field theories with metastable vacua in $3+1$ dimensions.

In the next section, we describe the Gross-Neveu model in detail. In
particular, we exhibit the phase structure of the theory in the presence of an
explicit chiral symmetry breaking term, and construct the exact bubble for this
theory. We then proceed to compare the exponential part of the activation rate
obtained from the exact bubble to that found from various commonly used
approximations, such as the Landau-Ginzburg approximation, amongst others.  We
next calculate the activation rate in the presence of finite fermion number.
Here we find that beyond a critical baryon number, the activation barriers
disappear completely.
Finally, section 5 contains our conclusions.

\section{The N-flavor Gross-Neveu Model}

We consider the Lagrangian density for the N-flavor
Gross-Neveu\cite{gn} model in $(1+1)$ dimensions:
\begin{equation}
{\cal L} = \bar\psi_{\alpha} (i\dslash + m_0) \psi_{\alpha} +
\frac{\lambda_0}{2N}  (\bar\psi_{\alpha}
  \psi_{\alpha})^2
\end{equation}
with $\alpha=1 \cdots N$ and a summation convention on $\alpha$ is used.
Introducing an auxiliary (Hubbard-Stratonovich) scalar field $\Delta_0$,
the Lagrangian density becomes
\begin{equation}
{\cal L} = \frac{-\Delta_0^2 N}{2 \lambda_0}+ \bar\psi_{\alpha}
(i \dslash +
m_0+\Delta_0) \psi_{\alpha} \label{Lagden}
\end{equation}

For $m_0=0$ the Gross-Neveu model has a {\it discrete} chiral symmetry:
\[\psi_{\alpha}\rightarrow i\gamma_{5}\psi_{\alpha} \; \; ; \; \;
\bar\psi_{\alpha} \rightarrow \bar\psi_{\alpha}i\gamma_{5}\; \; ; \; \;
\Delta_0 \rightarrow -\Delta_0 \]
It is this symmetry that will be spontaneously broken
(only discrete symmetries can be spontaneously broken in $1+1$ dimensions).
The explicit mass term $m_0$ introduces an explicit chiral symmetry breaking
term and will be responsible for a non-degenerate vacuum structure.

We should note that at tree level, the field $\Delta$ has {\em no} dynamics; it
is truly an auxiliary field. However, it acquires dynamical content by dint of
quantum effects arising from integrating out the fermions\cite{coleman}.

\subsection{The Gap Equation}

Let us first consider homogeneous configurations $\Delta(x)=\Delta$ and the
case of zero temperature.
In order to understand this case clearly, it is convenient to pass to the
Hamiltonian
\begin{equation}
H=\int \d x \left\{\frac{\Delta_0^2 N}{2 \lambda_0}+\psi^{\dagger}_{\alpha}
[-i\sigma_3 \frac{d}{dx}+\sigma_1 M_0]\psi_{\alpha}\right\}\label{hamil}
\end{equation}
with $M_0 = m_0+\Delta_0$ being the effective fermion mass, and the
$\sigma_i$ are the Pauli matrices.
The ground state is obtained by filling up the negative energy Dirac sea,
so that the ground state energy density is given by:
\begin{equation}
\frac{E}{NL}= \frac{\Delta_0^2}{2\lambda_0}-\frac{1}{\pi}\int_0^{\Lambda} \d k
\sqrt{k^2+M_0^2} \label{constener}
\end{equation}
where $L$ is the spatial length of the system and $\Lambda$  an upper
momentum cutoff. Dropping a term proportional to $\Lambda^2$ (normal ordering)
the energy density becomes
\begin{equation}
\frac{E}{NL}= \frac{\Delta_0^2}{2\lambda_0}-\frac{M_0^2}
{4\pi}-\frac{M_0^2}{2\pi}
\ln\left(\frac{2\Lambda}{|M_0|}\right) \label{ener}
\end{equation}
while the gap equation $(\partial E/\partial \Delta_0=0)$ becomes
\begin{equation}
\frac{\Delta_0}{\lambda_0}= \frac{M_0}{\pi}\ln\left
(\frac{2\Lambda}{|M_0|}\right)
\label{baregap}
\end{equation}
Clearly both the energy density and the gap equation need non-trivial
renormalizations. We find that the proper renormalization conditions are
\begin{eqnarray}
                M_0 & = & M_R = \Delta_R+m_R \label{Mren} \\
\frac{1}{\lambda_R} & = & \frac{1}{\lambda_0}-\frac{1}{\pi}\ln
\left(\frac{2\Lambda}{\kappa}\right) \label{lambdaren} \\
\frac{M_R}{\lambda_0}-\frac{m_0}{\lambda_0}
                    & = & \frac{M_R}{\pi}\ln\left(\frac{2\Lambda}{|M_R|}
                    \right)\label{relren} \\
\frac{m_0}{\lambda_0}
                    & = & \frac{m_R}{\lambda_R} \label{mren}
\end{eqnarray}
where $\kappa$ is an arbitrary renormalization scale.
The renormalized gap equation now becomes:
\begin{equation}
\frac{M_R}{\lambda_R}-\frac{m_R}{\lambda_R}=\frac{M_R}{\pi}\ln
\left(\frac{\kappa}{|M_R|}\right) \label{rengap}
\end{equation}

We see that for $m_R=0$ the gap equation generates dynamically  a scale
\begin{equation}
|\Delta_g| = \kappa e^{-\pi/\lambda_R} \label{dynamicalgap}
\end{equation}
This solution to the gap equation has the lowest energy (see below), thus
spontaneously breaking the discrete chiral symmetry.
It is convenient to absorb this dynamically generated scale
as the overall
scale in the problem and define all dimensionful quantities
in terms of this scale. We thus introduce
the following {\it dimensionless}
quantities
\begin{equation}
M     =  \frac{M_R}{|\Delta_g|}  \; \; ; \; \;
m   =  \frac{m_R}{|\Delta_g|} \label{etadelta}
\end{equation}
Discarding terms independent of $M$, the {\it dimensionless} energy
density
becomes
\begin{equation}
\frac{E}{N L |\Delta_g|^2} = \frac{M^2}{4\pi}[\ln(M^2)-1]-
\frac{M m}{\lambda_R} \label{etaener}
\end{equation}
and the gap equation becomes
\begin{equation}
\frac{M}{2\pi}\ln(M^2)= \frac{m}{\lambda_R} \label{newgap}.
\end{equation}
Clearly, for $m=0$ the solution $M = \pm  1$ gives the lowest energy,
thus the discrete chiral symmetry is indeed spontaneously broken.

The above energy density for a {\em constant} $\Delta_R$ configuration is
identified as the fully renormalized effective potential for the auxiliary
scalar field.

At finite temperature $\bar T$ we compute the free energy for the constant
configuration and thus obtain the finite temperature effective potential.
The free energy is given by
\begin{equation}
F[\Delta] = -\bar T\ln({\rm Tr\,} \exp(-H[\Delta]/\bar T))
\end{equation}
where the trace is over the fermionic degrees of freedom. Discarding
irrelevant terms, and introducing (along with $M,\ m$ above) the dimensionless
temperature $T = \bar T/|\Delta_g|$, we find the
dimensionless free energy per unit length to be (see Fig.~1)
\begin{eqnarray}
\Veff(M) \equiv \frac{F[M]}{N L |\Delta_g|^2} & = & \frac{M^2}{4\pi}
[\ln(M^2)-1]-\frac{M m}{\lambda_R}-\frac{2T^2}
{\pi}I(\frac{M}{T}) \label{freenercon} \\
I(y)                                  & \equiv &
 \int_0^{\infty} \d x \ln\left[ 1 + e^{-\sqrt{x^2+y^2}} \right]
 \label{integral}
\end{eqnarray}
A small-$y$ expansion of $I(y)$ is given by\cite{dj}
\Eq I(y) = {\pi^2\over12} + {y^2\over4} \left[ \ln(y/\pi) + \gamma - 1/2
  \right] - {7 \zeta(3) \over64\pi^2} y^4 + \ord{y^6} \Endl{iexp}
where $\gamma\approx 0.5772$ is Euler's constant and $\zeta(3)\approx 1.202$ is
the Riemann zeta function.

The extrema of the free energy are obtained from the gap equation:
\begin{equation}
\frac{M \ln M^2}{2\pi} - {2T\over\pi} I'\left(M \over
  T\right) = \frac{m}{\lambda_R} \label{gap}
\end{equation}
This equation can either have three solutions (corresponding to a local minimum
[metastable state] at $M=M_- < 0$, a maximum, and a global minimum
[stable state] at $M=M_+ > 0$ of the free energy), or just one solution
(minimum). The left hand side of (\ref{gap}) that determines the local
minimum at $M_-$ is plotted in Fig.~2 for various values of
$T$; we
see that solutions only exist for $m/\lambda_R \le 1/(e\pi) \approx 0.117$, and
for a given $m$ only over a certain range of $T$'s (\eg\ for
$m=0$, $T \le e^\gamma/\pi \approx 0.567$).

\subsection{The ``Bubble'' Configuration}

Static topological and non-topological semiclassical solutions in the
Gross-Neveu model have  been analyzed and systematically constructed by
Dashen, Hasslacher and Neveu (DHN)\cite{dhn} and Campbell and Liao\cite{cl}
using  inverse scattering techniques. These solutions were also studied
within the context of electron-phonon models of conjugate
polymers\cite{cb,fbc,bk,boy} and we refer the reader to these
references for a comprehensive analysis of these solutions.

For the case $m=0$ (explicit chiral symmetry) there are two
degenerate solutions to the gap equation.  In this case there
exist semiclassical topological soliton (kink) solutions and
non-topological bubble (polaron) solutions\cite{dhn,cl,cb}.
However when $m \neq 0$, the effective potential (or free energy density)
has one local and one global minimum, the degeneracy is lifted by the
explicit chiral symmetry breaking mass term and no soliton solutions
are available. However, by taking the results from the degenerate case,
it is straightforward to find that in the $m \neq 0$ case there
still exist bubble solutions for which the ``gap'' parameter $\Delta$ (and
therefore $M$) is spatially varying. These solutions (obtained via
inverse scattering methods\cite{dhn,cl,cb}) are given by
\begin{equation}
 M_{\pm}(x) = M_{\pm} - k_0
 \left\{ \tanh\left[k_0(x-x_0+y)\right] - \tanh\left[k_0(x-x_0-y)\right]
  \right\}\label{pol}
\end{equation}
where $M_{\pm}$ are the minima
solutions to the gap equation (\ref{gap}).  The parameters $k_0$ and $y$
are related by the ``integrability condition''\cite{dhn,cl,cb}
\begin{equation}
\tanh[2k_0y] = \frac{k_0}{M_{\pm}}. \label{integcon}
\end{equation}
Note that we may take $k_0 \geq 0$ since the sign of $k_0$ cancels on both
sides of the equation. Furthermore, notice that the solution
requires that $|y| \geq 1/|2M_{\pm}|$.  This is because the
bound states cannot be localized in distances smaller than the
Compton wavelength of the fermion.
The position of the center of mass of the bubble is determined by $x_0$.
This variable reflects the underlying translational invariance of the system.
Because of this invariance, the
energy (or free energy) is independent of this variable. For
notational simplicity we will  take it to be zero in the rest of
the analysis.

The fermion spectrum in presence of this non-topological semiclassical
configuration is also known exactly\cite{cl,cb,fbc,bk,boy}: N bound states
at energies
\begin{equation}
 \pm  w_{0 \pm} = \pm \sqrt{{M_{\pm}}^2-k_0^2}\;, \label{boundstateen}
\end{equation}
and positive and negative energy continuum states with energies
$\pm w_{k \pm} = \pm \sqrt{M_{\pm}^2+k^2}$ and phase-shifts
\begin{equation}
\delta(k) = 2\tan^{-1}\left(\frac{k_0}{k}\right). \label{phaseshift}
\end{equation}
with $k_0$ the solution of the integrability equation (\ref{integcon}). The
exact fermionic wave-functions may be found in reference\cite{cb}. A noteworthy
feature of these wave-functions is that the bound-state fermionic
wave-functions are localized at $x=\pm y$. It is known\cite{jcrb,jcshr} that in
the presence of a topological soliton (kink) there are fermionic zero modes
whose wave-functions are localized at the position of the kink. Since the
bubble configuration is reminiscent of a kink-antikink pair separated by a
distance $\approx 2|y|$, the localized fermion bound-states are the symmetric
and antisymmetric combination of the zero-mode wave-functions, split off in
energy because of the non-zero overlap of these wave-functions.

The above configuration, being similar to a kink-antikink pair separated by a
distance $2|y|$, can clearly be identified with a ``bubble''. The value of $y$
(the remaining unspecified parameter) may be obtained by extremizing the energy
of this configuration as a function of this parameter.  For the moment we will
leave this parameter free and treat it as a ``collective coordinate''.

Before proceeding to the numerical analysis of the bubble energy, it proves
illuminating to understand some features of this configuration.

Asymptotically this solution reaches $M_{\pm}$. For the global minimum $M_+ >0$
and the integrability condition (\ref{integcon}) implies that $y >0$. In this
case, in the region $-y < x < y$, the bubble samples a region where the free
energy (or energy) density is {\it higher} than that of the global minimum (for
large $y$, probing the metastable state). Thus, for large $y$, the energy as a
function of $y$ will {\it grow} linearly (because the difference of the tanh's
is roughly constant in this region). On the other hand, if the bubble solution
has its asymptotics in the {\it metastable} minimum for which $M_- <0$, then
$y<0$ and the bubble configuration is probing a region ($y<x<-y$) in which the
free energy (or energy) density is {\it lower}, thus {\it gaining} volume
energy.  In this case for large $y$ the energy of the bubble configuration will
{\it diminish} (linearly for large $|y|$). Certainly for small $y$ the spatial
variations of the bubble configuration will raise the energy, and since for
large $y$ the energy will diminish, there has to be a maximum of the energy as
a function of $y$.  This maximum, the critical bubble, thus corresponds to a
``sphaleron'' in the sense of Manton and Samols\cite{mansa}.  The maximum of
the energy as a function of the collective coordinate $y$ corresponds to a
saddle point in functional space. The unstable coordinate corresponds to
$\delta y = y-y^*$ where $y^*$ corresponds to the maximum of the energy. The
collective coordinate $x_0$ corresponding to the center of mass of the bubble
is a flat direction in functional space, along which the energy is constant as
a consequence of translational invariance.
This configuration is an {\it exact saddle point} of the
full effective action (functional of $\Delta(x)$) obtained
after integrating out the fermions
\begin{equation}
S_{eff} = N \int d^2x \left\{-\frac{\Delta_0^2}{2 \lambda_0}
\right\}
-i N\, {\rm Tr} \ln [i \dslash +m_0+\Delta_0]. \label{effectiveaction}
\end{equation}
Finite temperature enters in the trace via the boundary conditions
on the fermionic fields.
In references (\cite{dhn,cl,cb})
the effective action is written in terms of the scattering
data (bound state energies and normalization and phase shifts) and
the bag configuration is found by extremizing the full effective
action with respect to these parameters. (This is the essence
of the inverse scattering method). In the large N limit
the functional integral is determined completely by the
 saddle points. However, even for finite N, the critical
droplet configuration is an {\it exact} saddle point of the
effective action.

Hereafter we will choose (making use of dimensional transmutation) the
renormalization scale $\kappa$ to be the value (analogous to $\Lambda_{\rm
QCD}$) for which $\lambda_R=1$.  In units of the dynamically generated scale,
the renormalized free energy difference between the metastable bubble
configuration and that of the constant ``gap'' configuration in the metastable
state is
\begin{eqnarray}
\frac{F}{N} &=& \frac{1}{2}  \int\d x \left(\Delta^2(x)-
\Delta_-^2 \right)  - w_{0-} -
    \int_0^{\kappa/2} {\d k \over\pi}{\d \delta\over\d k} w_{k-} +
    {2 k_0\over\pi}
    \nonumber\\
            &-& 2T\ln\left[ 1 + e^{-w_{0-}/T} \right] - 2T \int_0^\infty
            {\d k\over \pi}
    {\d \delta \over\d k} \ln\left[ 1 + e^{-w_{k-}/T} \right] \label{F1}
\end{eqnarray}
The phase-shift $\delta(k)$ is given by (\ref{phaseshift}), the bound state
energy $w_{0-}$ is given by (\ref{boundstateen}), and $w_{k-}$ are the
continuum energies.  $F/N$ then simplifies to
\begin{eqnarray}
 \frac{F}{N}
   &=& {k_0\over\pi} \left[ 4\pi m y + 2 - \ln M_-^2
    - {2w_{0-}\over k_0} \tan^{-1}\left( k_0\over w_{0-} \right) \right]
    \nonumber\\
   &-& 2T\ln\left( 1 + e^{-w_{0-}/T} \right) + {4T k_0\over\pi} \int_0^\infty
   {\d k \over k^2+k_0^2} \ln\left( 1 + e^{-w_{k-}/T} \right)
  \label{F2}
\end{eqnarray}
The first term inside the square brackets shows that the energy diminishes for
large $|y|$ (recall $y<0$), consistent with the argument presented above. This
is the classical ``volume'' energy contribution and arises primarily from the
first (integral) term in (\ref{F1}).  The rest of the bracketed expression is
the $T=0$ quantum contribution from the fermionic degrees of freedom, and
saturates at a constant value for large $|y|$.  The remaining terms give the
thermal contribution.

We want to use the above results to calculate the activation rate between the
metastable and the true vacua. Our procedure will be as follows. We have an
{\em exact} solution to the full, quantum equations of motion. To calculate the
activation rate we can use the path integral formalism\cite{linde}; what we do
is to saturate the path integral with the bubble configuration found above and
evaluate the rate semiclassically. This is just the standard procedure that is
commonly used. Thus, when we say we are calculating the ``exact'' activation
rate, we mean that we are using the exact solution to the full equations of
motion to perform the semiclassical calculation of the rate. In the large $N$
approximation, our calculation would become exact in {\em all} senses of the
word.


In the large N limit the gaussian fluctuations around the saddle point
configuration yield a contribution of order $(1/N)$ and to leading order the
prefactor is $1$. However the free energy $F= N (F/N)$, with $F/N$ given by
equation (\ref{F2}), and the rate formally vanishes in this limit.

For finite N we should take into account the corrections from the gaussian
fluctuations around the saddle point configuration, that is the prefactor, this
is an extremely difficult task in this case.

Therefore, we only concentrate on the exponential since it gives the leading
behavior and in this case it contains the {\em quantum corrections} associated
with the fermionic loops, in particular in the case under consideration, the
contribution to the exponential is {\em solely} arising from quantum
corrections, since at tree level the auxiliary field has no dynamics.

Having clarified the nature of the approximations involved, we now set $N=1$.


Given $m$ and $T$, then, we can plot $F$ as a function of $y$ (Fig.~3), and the
maximum is $F^C_c$.  The superscript ``C'' indicates that this is the
``Correct'' method of calculating the critical bubble free energy.  Up to a
prefactor, the (semiclassical) activation rate is then $\Gamma^C(T) =
\exp(-F^C_c/T)$.  For $m=0.02$, $\Gamma^C(T)$ is plotted as a solid curve
in Figs.~5--8.

\section{Some Approximation Schemes}

Now we turn to the main point of this work, which is to compare the {\em exact}
critical free energy in the GN model to that obtained by means of various
approximations.

There are a variety of ways to approximate the free energy that enters into the
activation rate. The class of schemes we consider involve approximating the
kinetic term in various ways (note that the theory written in terms of the
auxiliary field $\Delta$ has no canonical kinetic term at tree level) as well
as approximating the potential term. Some of these schemes will have
counterparts in the $3+1$ dimensional calculations of activation rates in
scalar fields, while others are more specific to the Gross-Neveu model.

\subsection{The Landau-Ginzburg Approximation}

The BCS theory of superconductivity contains a four-fermi interaction which
controls the formation of Cooper pairs. A complex auxiliary field $\phi$ can
be introduced to rewrite this four fermi interaction as a Yukawa type
coupling. This field then serves as an order parameter for the superconducting
state. Its dynamics can be determined by integrating out the fermions and
constructing the effective action for $\phi$. To understand the nature of the
phase transition in this case, it is useful to perform a derivative expansion
of the effective action near the critical temperature. This is the
Landau-Ginzburg expansion\cite{sakita}. We can perform the same procedure for
the Gross-Neveu model.

The Landau-Ginzburg effective free energy  is obtained as a consistent
expansion in small order parameter ($M$) and small gradients, and is valid near
the critical temperature and for small explicit symmetry breaking fields (small
$m$). In this approximation (method ``L'') the effective free energy density is
written up to terms of order $(\partial M(x)/ \partial x)^2$ in the derivative
expansion and up to order $M^4(x)$ in the non-derivative (potential) terms. The
calculation of the lowest gradient term is obtained via a Feynman diagram
expansion to one loop using the imaginary time formulation of finite
temperature field theory. It is carried out by writing $M(x)=M+v(x)$ with $M$
the {\it homogeneous} configuration, and $v(x)$ (the small departure from the
homogeneous configuration) taken to be a perturbation.  Due to the Yukawa
coupling of $M(x)$ to the fermions, $M$ will be the mass of the fermions in the
loop.

The coefficient of the gradient term is the bracketed expression in:
\Eq {i\over2} \, \raisebox{-7pt}{\begin{picture}(40,20)
   \put(0,10){\line(1,0){4}} \put(6,10){\line(1,0){4}} \put(20,10){\circle{20}}
   \put(30,10){\line(1,0){4}} \put(36,10){\line(1,0){4}} \end{picture}}
  = p^2 v(p) v(-p) \left[ {1\over24\pi M^2} - {1\over 12\pi M T}
   I''' \left( M\over T \right) \right]  \Endl{lg1}
where $I(y)$ was defined in \puteq{integral}.  The diagram shows a fermion loop
attached (with coupling $v$) to two (truncated) scalar legs carrying spatial
momentum $p$. The field $v(x)$ is taken to be independent of the Matsubara
frequencies, as well as slowly varying in space, in keeping with the philosophy
of the derivative expansion.  In the $T\to 0$ limit the bracketed expression is
$1/(24\pi M^2)$, while for $T\gg M$ it is [using the expansion of \puteq{iexp}]
$7 \zeta(3)/(32 \pi^3 T^2)$.  We have verified this calculation by means of
Chan's method of obtaining the derivative expansion\cite{chan}.

Using the quartic approximation (for $T\gg M$) to the finite $T$ effective
potential of
\puteq{freenercon} (see Fig.~4) yields the Landau-Ginzburg
free energy density (per flavor):
\Eq {\cal F}^{\rm LG} = {7 \zeta(3)\over 32 \pi^3 T^2} (\partial_x
  M)^2 -M\ m + {M^2\over2\pi} \left[\ln(\pi
  T) - \gamma\right] + {7 \zeta(3) \over 32\pi^3}
{M^4\over
  T^2} \Endl{LLG}
This free energy density can be extremized with respect to
$M(x)$,
and it is found that the extremum inhomogeneous configuration
is a  bubble solution\cite{ccar} which is of the same form as the true
extremal bubble, \puteq{pol},
with $k_0$ fulfilling exactly the {\it same} integrability
condition as in \puteq{integcon}.  However, in this case,
$M_-$ is the
local minimum of the Landau-Ginzburg potential, and for this extremal
bubble,
\Eq k_0^2 = \frac{3 M_-^2}{2} + {4\pi^2 T^2 \over 7\zeta(3)}
  \ln\left( \pi T e^{-\gamma} \right) \End
With $t \equiv k_0/|M_-|$, the total  free energy of this
inhomogeneous configuration in the
Landau-Ginzburg approximation is
\Eq F^L_c = {7 \zeta(3)\over 4 \pi^3 T^2} k_0^3 \left[ {-4\over3} +
  {2\over t^2} + {1-t^2 \over t^3}
  \ln\left( 1-t \over 1+t \right) \right] \Endl{FLc}
The exponential part of the activation rate, $\Gamma^L(T) = \exp(-F^L_c/T)$, is
plotted as a dashed curve in Fig.~5, along with the correct result
$\Gamma^C(T)$; from this graph we see that method ``L'' overestimates the rate
up until the critical temperature where they both become equal to unity (the
critical temperatures are slightly different).  This feature will recur in all
the approximation schemes we use here, except for the last (zero $T$ bubble).

\subsection{Landau-Ginzburg With $T=0$ Gradients}

In method ``G'', the same Landau-Ginzburg potential is used, but only the $T=0$
gradient term is taken, \ie\ the coefficient in front of the derivative term is
$1/(24 \pi M_-^2)$.  The critical bubble free energy is then given by
\begin{equation}
F^G_c = \frac{F^L_c}{R}, \qquad R \equiv {\sqrt{24 \pi M_-^2}} {\sqrt{\frac{7
  \zeta(3)} {32 \pi^3 T^2}}}
\end{equation}
where $F^L_c$ was given in \puteq{FLc}.  $\Gamma^G(T) = \exp(-F^G_c/T)$ is
plotted as a dashed curve in Fig.~6.

This method is closer in spirit to the one used in the standard
calculations\cite{linde} of finite temperature activation rates for scalar
field theories with canonical kinetic terms. In these calculations, the
canonical kinetic term is combined with the finite-$T$ effective potential to
give the free energy relevant to the activation rate.

\subsection{The Effective Potential Approximation}

In the effective potential approxmation (method ``P''), we define
\Eq \Fpot = \int\d x \left[ \Veff(M(x)) - \Veff(M_-) \right] \Endl{Fpot}
where $\Veff$ was defined in \puteq{freenercon}.  Among configurations of the
form \puteq{pol} [and satisfying the integrability condition \puteq{integcon}],
parameterized by the half-width $y$, we choose $y$ to extremize $\Fpot$, and
the resulting $\Fpot$ is our approximation to the critical free energy $F^P_c$.
$\Gamma^P(T) = \exp(-F^P_c/T)$ is plotted as a dashed curve in Fig.~7.

Since {\it classically\/} there is no kinetic term in our Lagrangian,
\puteq{Lagden}, this method is tantamount to {\it ignoring derivative
corrections\/} (both quantum and thermal).  In real-world calculations, such as
that of the free energy of a critical bubble or the mass of a sphaleron,
derivative corrections are often ignored \cite{clee}; the finite-$T$ effective
potential is combined with the classical (canonical) kinetic term, and the
resulting approximate action is extremized.

For example, the sphaleron mass at $T=0$ is known to be \cite{klink}
\Eq M_{\rm SP} = {4\pi v B\over g}, \quad (B = 1.5-2.7) \End
where $v$ is the minimum of the Higgs potential.  The sphaleron mass at finite
$T$ is then approximated by the same formula but with $v\to v(T)$, the minimum
of the Higgs effective potential [and also $g\to g(T)$].  Derivative
corrections to the action have been ignored.

\subsection{The Zero-$T$ Bubble}

In the zero-T approximation (method ``Z''), the approximation scheme is as
follows. We start by finding the $T=0$ critical bubble, a configuration of the
form \puteq{pol} and \puteq{integcon}, with $M_-$ the minimum of {\em zero
temperature} effective potential, $\Veff(T\!=\!0)$, choosing $y$ to extremize
the $T=0$ free energy.  Then we plug this value of $y$ into the {\em
finite}-$T$ formula for the free energy, \puteq{F2}.  $\Gamma^Z(T) =
\exp(-F^Z_c/T)$ is plotted as a dashed curve in Fig.~8.

This is analogous to the method used for critical bubbles in \cite{clee}.
There the bubble configuration used was the extremum of the classical, $T=0$
action. The (1-loop) finite-$T$ free energy of this bubble was then calculated
(exactly, including ``derivative corrections'') by numerically evaluating the
determinant factor.

The purpose of ref.~\cite{clee} was to measure ``derivative corrections'' by
comparing the Zero-$T$ calculation (which includes derivative corrections) to
the effective potential calculation (which does not).  This effort was hampered
by the fact that the Zero-$T$ calculation is itself an approximation (only good
at lower temperatures).  In the Gross-Neveu model of this paper, we can compare
both methods to the true result.

\section{Finite Fermion Number}

If sphaleron configurations are relevant in baryon number
violating processes, an important question to pose is how
the presence of a finite baryon number modifies the field
profile and the free
energy barrier of the sphaleron configuration.
Within the model under consideration we can provide some
partial answers to these questions. In this model we identify
baryon number with fermion number. In order to consider a
finite baryon number {\it density} at zero temperature
 we would have to allow for the case in which the N positive
energy fermion bound states are occupied, and the rest of the
baryons to be in the positive energy continuum up to a
Fermi energy (or chemical potential). Because the phase shifts
of the continuum states depend on the bubble
 ``collective coordinate'' $y$, to obtain the bubble profile
for arbitrary $y$, the Fermi energy will have to be adjusted
such as to provide the constant baryon density (a constraint
on the system).

At finite temperature we would have to introduce a chemical
potential and work in the grand canonical ensemble, finally
fixing the chemical potential (as a function of temperature
and bubble size) to give the fixed baryon number density
(on average). Both situations are extremely hard to implement
 and at the present time lie beyond our capabilities.

We will content ourselves with considering the somewhat more
restricted scenario in which we have a {\it fixed baryon
number} (B), $0< B \leq N$ and the case of zero temperature. The reason
that this case is somewhat simpler is that we can accommodate
the $B$ baryons in the available N positive energy bound
states. This corresponds to the ground state configuration with
a finite (positive) baryon number $B \leq N$ (for a negative
baryon number we would have to deplete B {\it negative} energy
bound states).

Following the steps leading to (\ref{F2}) at $T=0$ we find
that the free energy becomes
\begin{equation}
\frac{F^B}{N} = \frac{F^0}{N} + \frac{B}{N}\omega_0
\end{equation}
with $F^0$ given by equation (\ref{F2}) with $T=0$. It is
convenient to parametrize
\begin{eqnarray}
k_0  & = & |M_-| \sin(\theta) \nonumber \\
\omega_0
     & = & |M_-| \cos(\theta)  \\
y(\theta)
     & = & \frac{1}{4 \sin(\theta)}\ln\left
[\frac{1+\sin(\theta)}{1-\sin(\theta)} \right] \nonumber \\
  0  & \leq
         & \theta \leq \pi /2 \nonumber
\end{eqnarray}
Since $y(\theta)$ is a monotonically increasing function
of $\theta$ it proves more convenient to locate the size of
the critical bubble by extremizing the free
energy with respect to $\theta$. The extremum condition
leads to
\begin{equation}
-\frac{m}{|M_-|}\tan(\theta)+\frac{\theta}{\pi} = \frac{B}{2N}
\label{sphalcond}
\end{equation}
As we have analyzed in section two, the metastable minimum
is available only for $\pi m/|M_-| < 1$.
Thus we find that for $B \neq 0$ there are two {\it non-trivial}
extrema of the energy functional. Clearly one corresponds to
a minimum and the other to the maximum, i.e. the critical bubble.
The free energy barrier between the minimum and the maximum
becomes smaller as $B$ is increased, finally disappearing at
a maximum baryon number given by $B_{max}$ where $B_{max}$ is found using
$\theta_{max} = \sin^{-1}\left( \sqrt{1-\pi \frac{m}{|M_-|}} \right)$ in
\puteq{sphalcond}.

For $B > B_{max}$ the barrier disappears altogether and there
is no longer a bubble solution. This is indeed a remarkable
result if it persists in three dimensions.
The physics of this phenomenon is clear. By filling up the
allowed positive energy bound states the Pauli pressure (a
consequence of the Pauli exclusion principle) increases. For
the minimum which appears at a {\it smaller} bubble radius this
pressure is larger than that for the maximum, which appears for
a larger bubble radius.
 Although extrapolation
to three dimensions is not warranted, we would expect
fermionic bound states in the lowest ($J= 1/2$) partial wave,
because in this partial wave the problem is essentially one-dimensional
(save for the fact that the wave function must satisfy proper
boundary conditions at the origin).
If the effect of these fermionic bound states localized at the
wall of the bubble is consistent with the behavior just found
in this $1+1$ dimensional model, it would imply that the
current bounds on sphaleron transitions may be a gross
{\em underestimate} for the rate. Clearly we cannot conclude that this
happens in the more realistic three dimensional scenario, but
we believe that this observation may be worthy of consideration
there as previous estimates did not take into account the
fermionic back-reaction onto the sphaleron configuration and
their effect on the activation barriers.

\section{Results and Conclusions}

The amazing thing about the GN model is that one can calculate the {\em exact}
critical bubble configuration of the theory and hence we can deduce the true
(modulo our statements about the semiclassical evaluation of the relevant path
integral) value of the activation rate (or at least its exponential part, which
typically is the most important part of the rate).

$\Gamma(T) = e^{-F_c/T}$ calculated by our several methods (one exact and four
approximations) are plotted in Figs.~5--8 (for $m=0.02$ and $N=1$).  The
bubbles $M(x)$ (for $m=0.02$, $N=1$, and $T=0.35$) used in these methods are
plotted in Fig.~9.

We see that the Landau-Ginzburg method (``L'') and the potential method (``P'')
converge nicely to the correct result (``C'') at high temperatures.  They both
correctly predict $\Gamma(T_c)=1$, though their respective critical
temperatures vary slightly from the correct $T_c$; we would only expect exact
agreement if the phase transition were second order (the limit $m\to 0$).  The
zero-T method (``Z'') is a poor approximation at higher temperatures, but is
better at lower $T$'s.  Curiously, all the approximation schemes tend to
overestimate the rate.

Finally, we have to address the question of what this means for realistic
field theories in $3+1$ dimensions. Since the GN model was so special in that
we could find the exact bubbles, etc., one might be tempted to think that
our results should only hold for this model. While we have no evidence that
this is not the case, it should be recalled that the GN model serves as a
very good testing ground for issues such as asymptotic freedom, which does
in fact occur in $3+1$ field theories.
It is also the inspiration for Nambu-Jona-Lasinio low energy
effective models for strong interactions.
 Thus, our results may have a wider range
of applicability to theories in higher dimensions. In particular, if the
various sphaleron configurations in higher dimensional theories can sustain
fermionic bound states, then we expect that our discussion of the activation
rate in the presence of finite fermion number should apply. This would be of
great importance since the reduction of the barrier would have significant
consequences for the activation rate.

At any rate, what our results {\em do} show is that the standard
methods of calculating the activation rate need not be all that close to the
exact value; if nothing else, our calculation can serve as a warning sign to
those who might believe their approximations overly much.

\vskip.4in plus.2in minus.1in
\leftline{\Large\bf Acknowledgements} \bigskip
DB and DSL were partially supported by NSF Grant \# PHY-9302534.  DSL was
also supported by a Mellon Fellowship.  DEB and RH were partially supported
by the U.S. Dept. of Energy under Contract DE-FG02-91-ER40682.


\def\ap#1{{Ann.\ Phys.} {#1}}
\def\pl#1{{Phys.\ Lett.} {#1}}
\def\np#1{{Nucl.\ Phys.} {#1}}
\def\pr#1{{Phys.\ Rev.} {#1}}
\def\prl#1{{Phys.\ Rev.\ Lett.} {#1}}
\def\prog#1{{Prog.\ Theor.\ Phys.} {#1}}
\def\rbf#1{{Rev.\ Bras.\ F\'is.} {#1}}
\def\zp#1{{Z.\ Phys.} {#1}}
\def\jcp#1{{J.\ Chem.\ Phys.} {#1}}
\def\yf#1{{Yad.\ Fiz.} {#1}}
\def\sjnp#1{{Sov.\ J.\ Nucl.\ Phys.} {#1}}
\def\ibid{{ibid.\ }}


\newpage
\INSERTFIG{5.5}{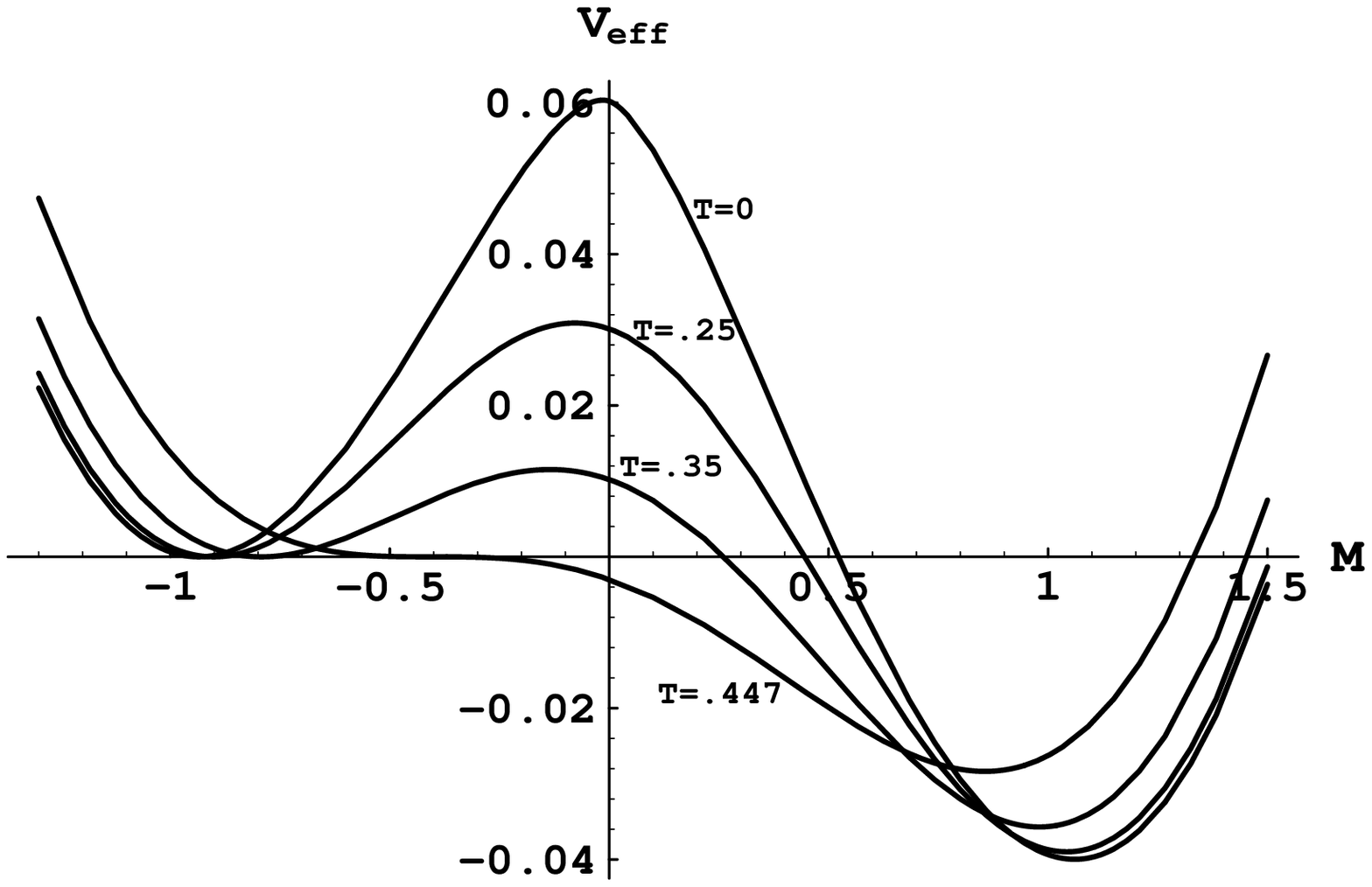}{Fig.~1: $\Veff(M)$ for $m=.02$ and
  $T=\{0,\,.25,\,.35,\,.447\}$.}
\vskip.3in
\INSERTFIG{5.5}{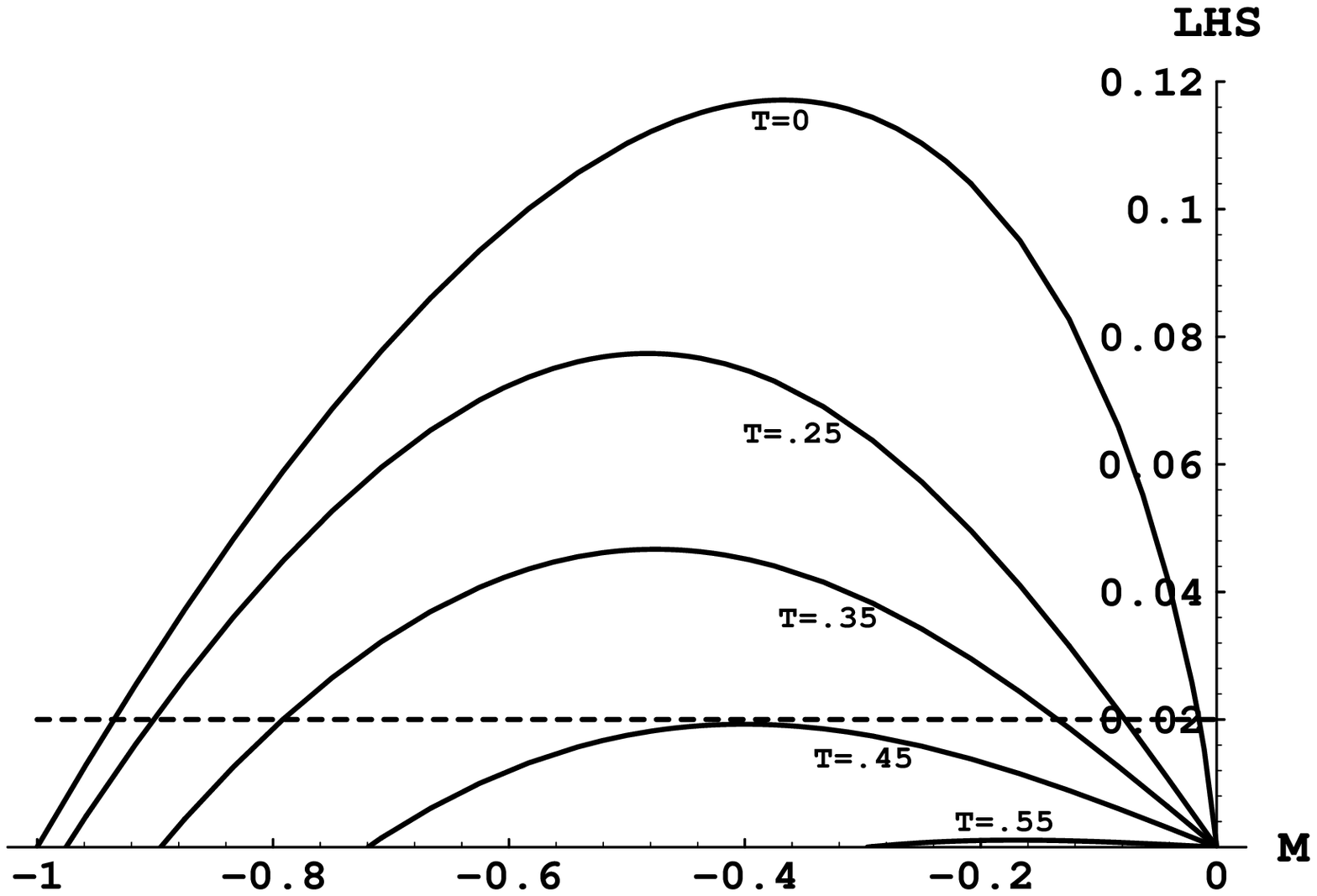}{Fig.~2: Left-hand side (LHS) of the gap equation at
  $T=\{0,\,.25,\,.35,\,.45,\,.55\}$.\\Dashed line is $m/\lambda_R=.02$, for
  which the critical temperature is $T_c = .447$.}

\figskip
\INSERTFIG{5.5}{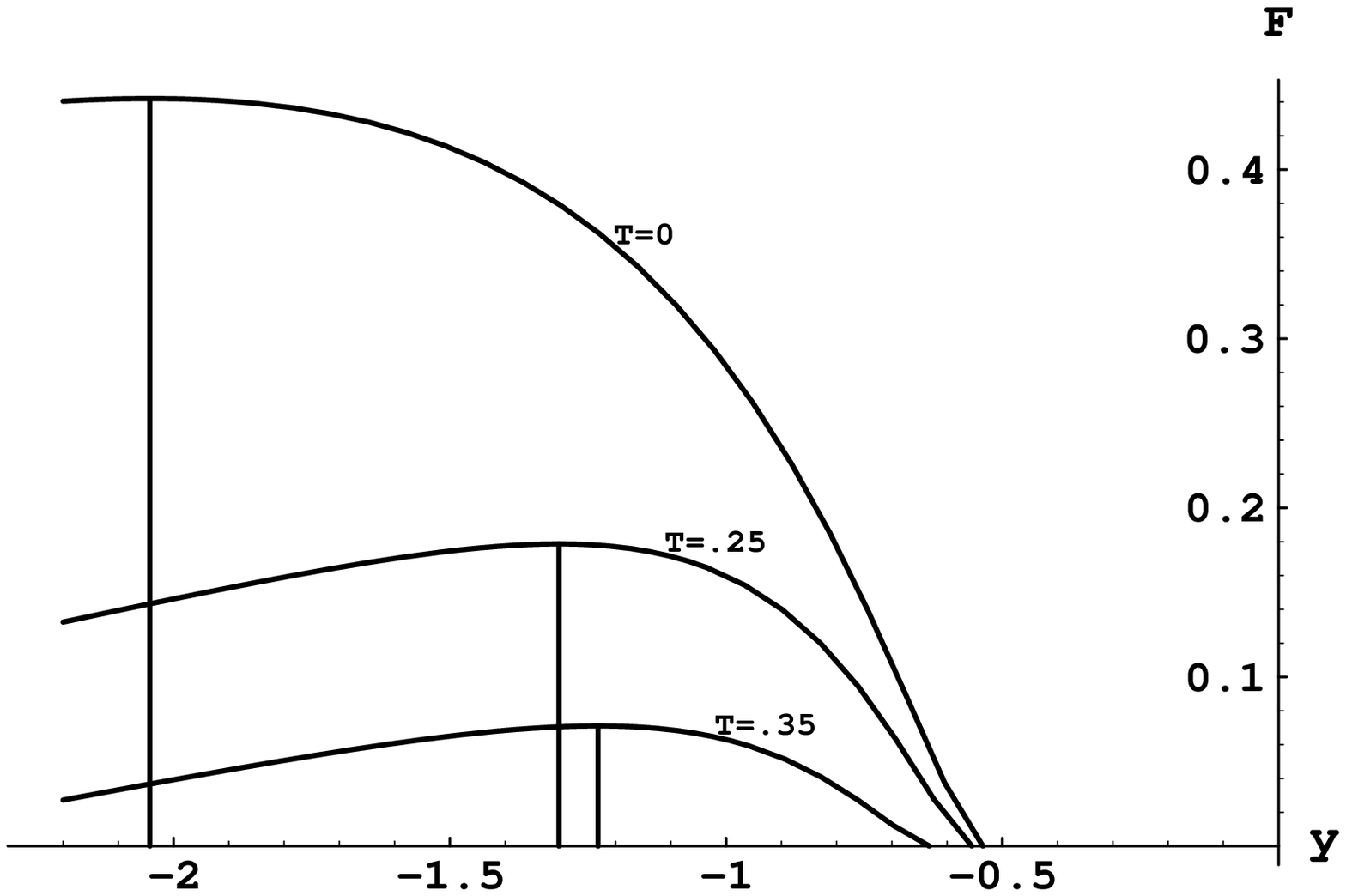}{Fig.~3: Bubble free energy $F(y)$ as a function of
  the half-width $y$, for $m=.02$\\and $T=\{0,\,.25,\,.35\}$.  The maximum
  (for a given $T$) is the critical free energy $F_c$.}
\vskip.3in
\INSERTFIG{5.5}{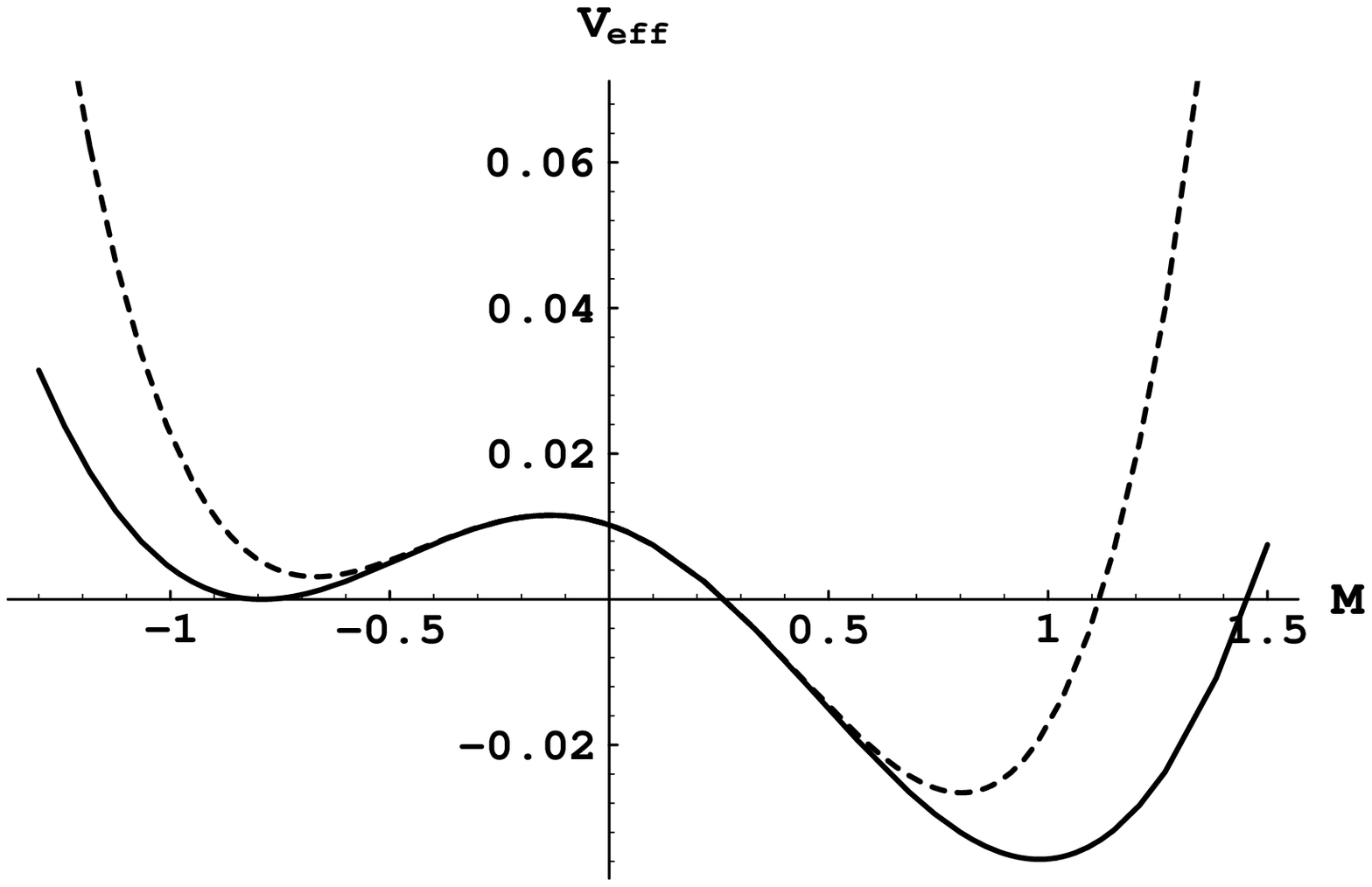}{Fig.~4: True $\Veff(M)$ (solid) and the
  Landau-Ginzburg\\approximation (dashed), for $m=.02$ and $T=.35$.}

\figskip
\INSERTFIG{5.5}{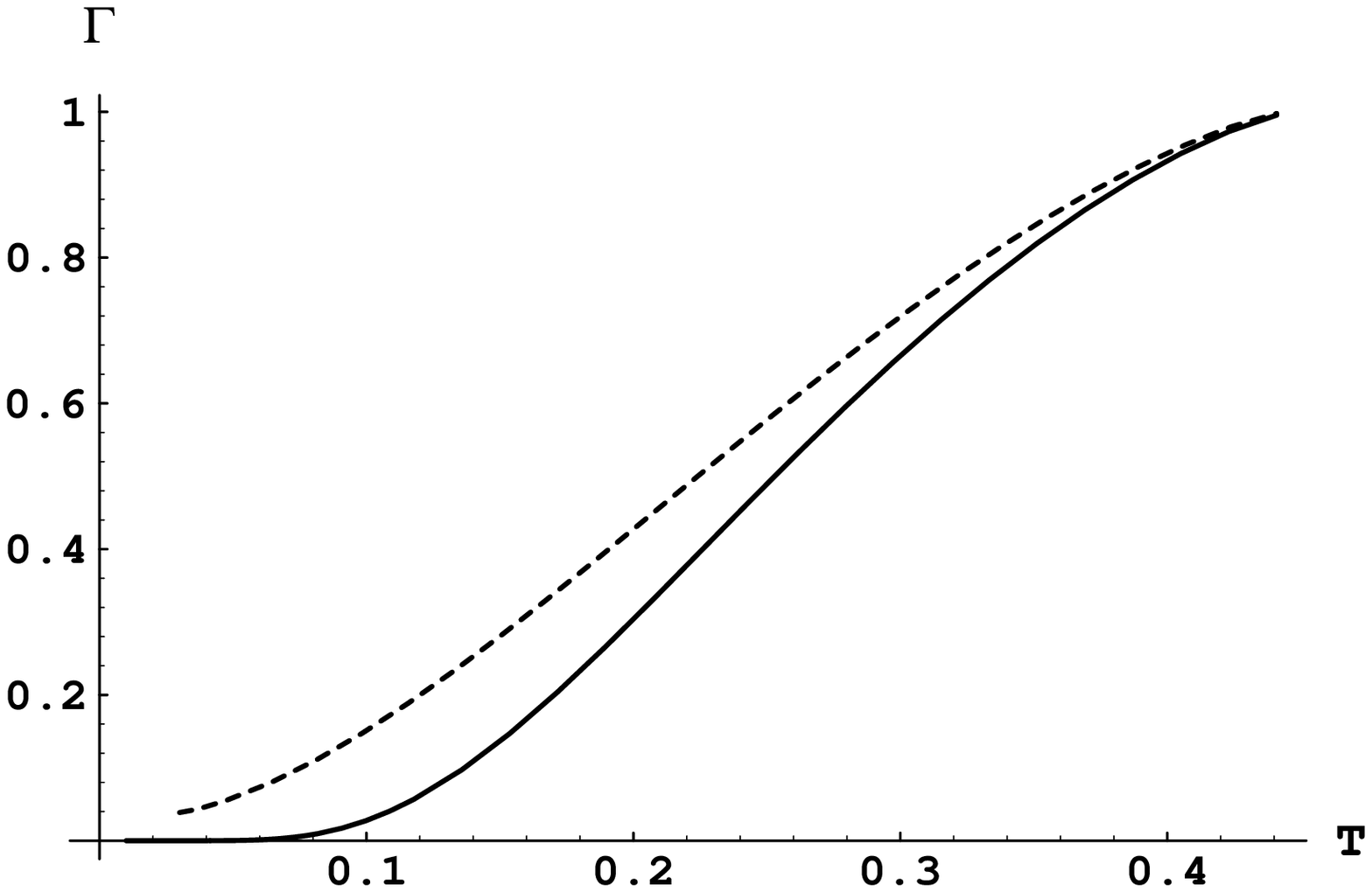}{Fig.~5: $\Gamma(T) = e^{-F_c(T)/T}$ for $m=.02$:
  solid = C (Correct),\\dashed = L (Landau-Ginzburg).}
\vskip.3in
\INSERTFIG{5.5}{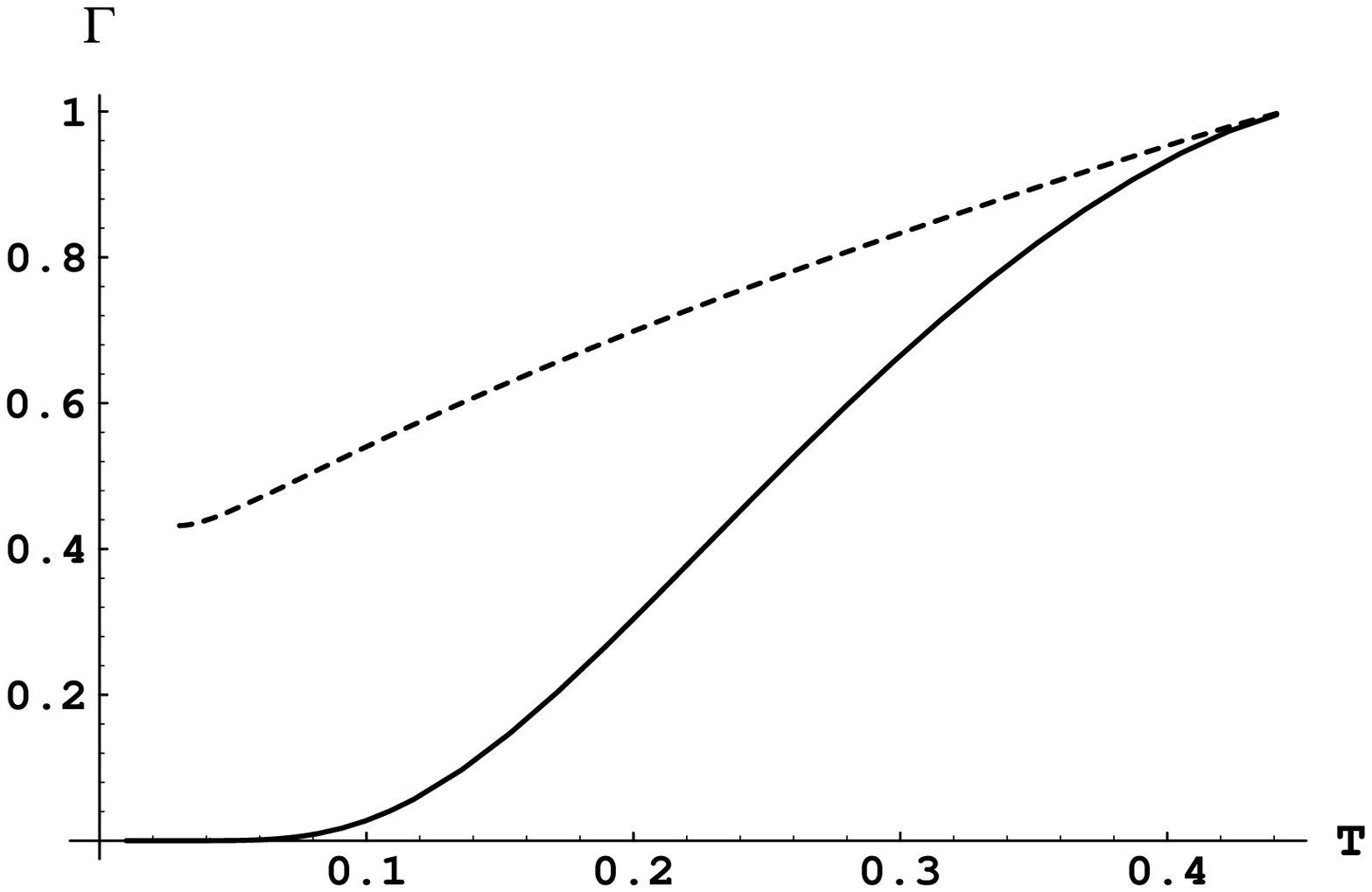}{Fig.~6: $\Gamma(T) = e^{-F_c(T)/T}$ for $m=.02$:
  solid = C (Correct),\\dashed = G (Landau-Ginzburg With $T=0$ Gradients).}

\figskip
\INSERTFIG{5.5}{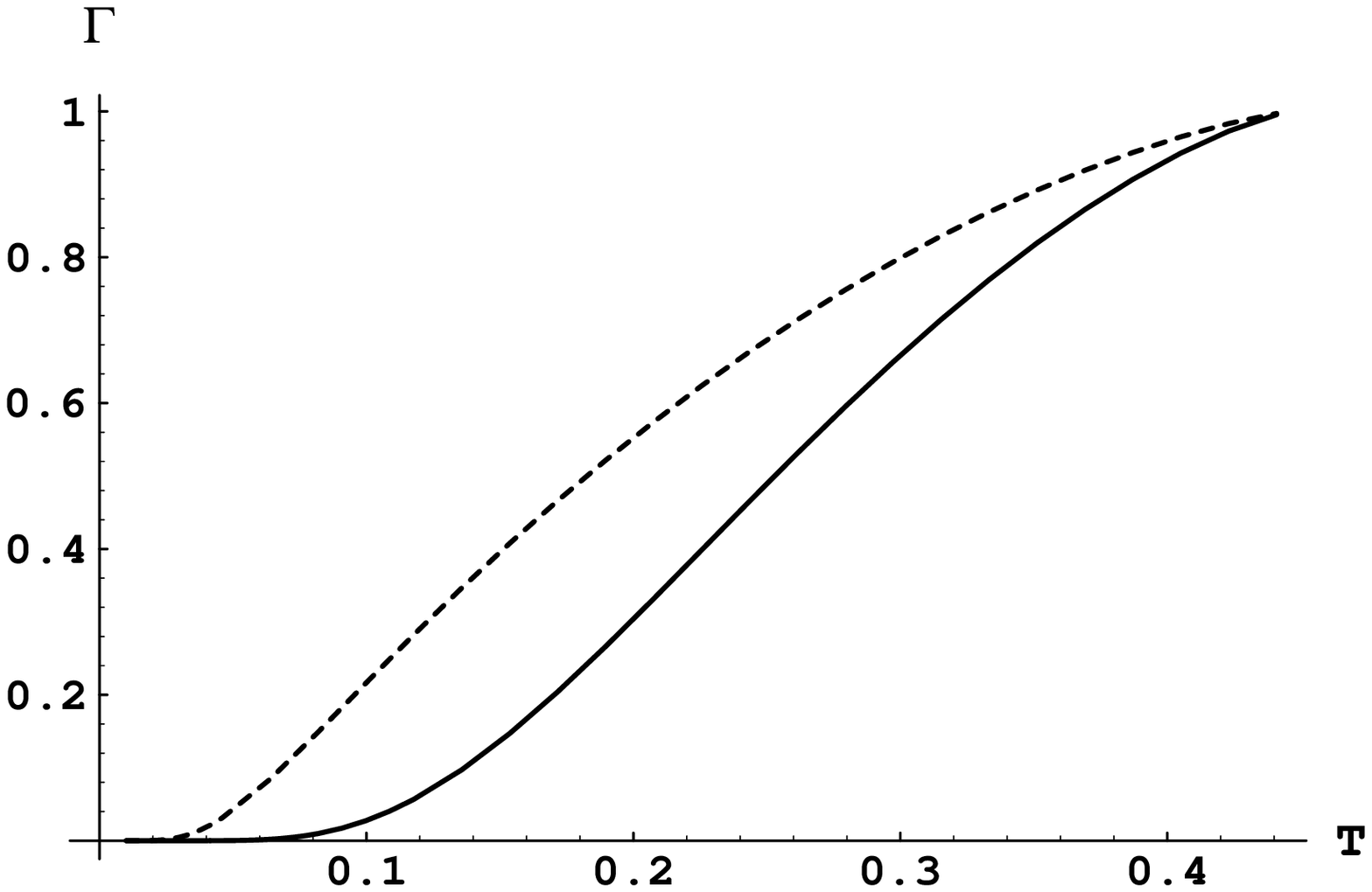}{Fig.~7: $\Gamma(T) = e^{-F_c(T)/T}$ for $m=.02$:
  solid = C (Correct), dashed = P (Potential).}
\vskip.3in
\INSERTFIG{5.5}{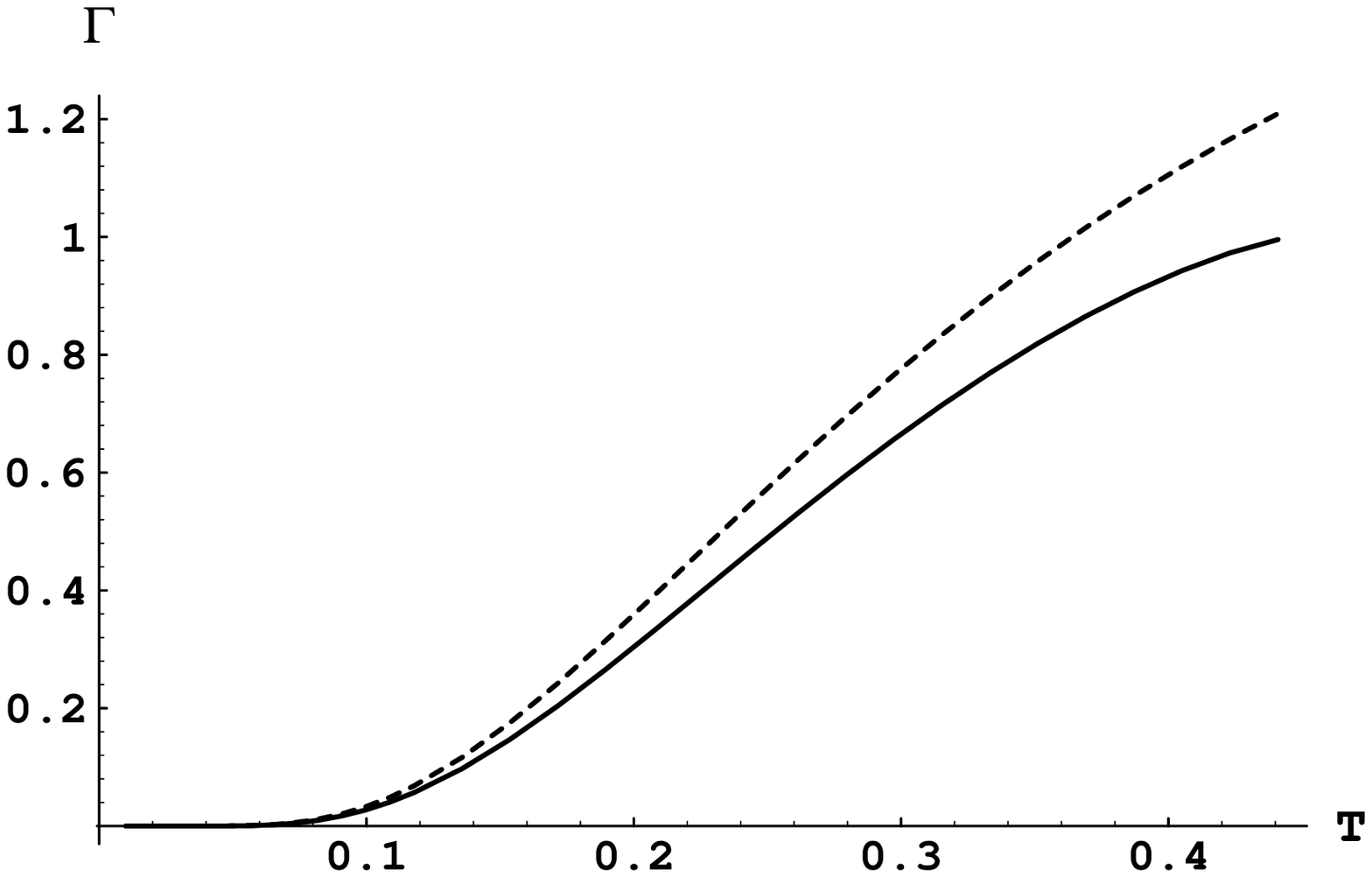}{Fig.~8: $\Gamma(T) = e^{-F_c(T)/T}$ for $m=.02$:
  solid = C (Correct), dashed = Z ($T=0$ bubble).}

\figskip
\INSERTFIG{6.4}{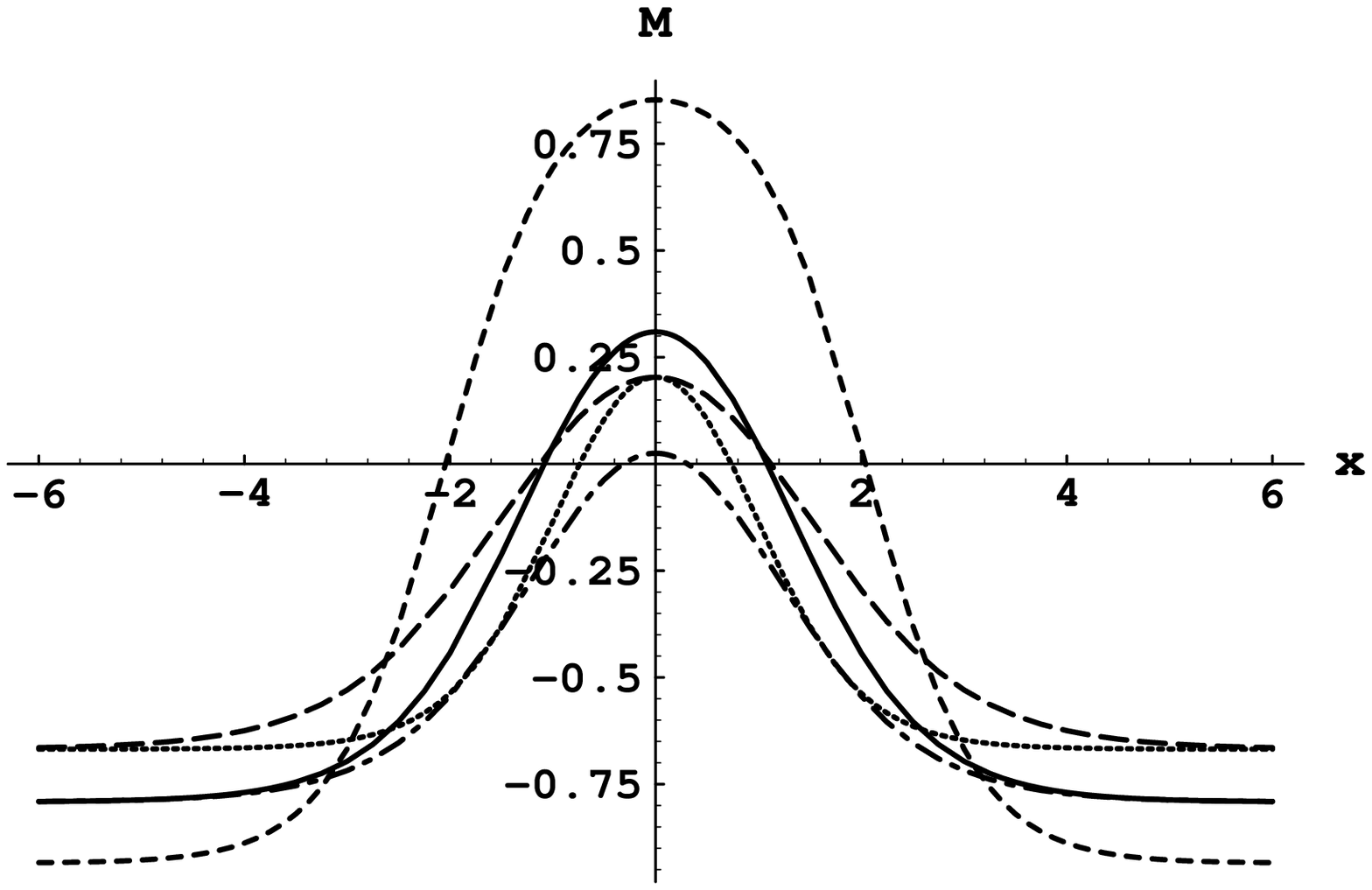}{Fig.~9: Critical bubbles $M(x)$ for $m=.02$
  and $T=.35$, five methods:\\solid = C (Correct), long-dash = L
  (Landau-Ginzburg),\\dots = G (Landau-Ginzburg With $T=0$ Gradients),
  \\dot-dash = P (Potential), short-dash = Z ($T=0$ bubble).}

\end{document}